\documentclass[aps,prb,floatfix, twocolumn,superscriptaddress]{revtex4-1}

\usepackage{graphicx}
\usepackage[usenames]{color}
\usepackage{psfrag}
\usepackage{amssymb}
\usepackage{eso-pic}
\usepackage[dvips,letterpaper]{geometry} 
\usepackage{amsmath}

\begin{document}

%
%
\def\oti{{\otimes}}
\def\lb{ \left[ }
\def\rb{ \right]  }
\def\tilde{\widetilde}
\def\bar{\overline}
\def\hat{\widehat}
\def\*{\star}
\def\[{\left[}
\def\]{\right]}
\def\({\left(}		\def\BL{\Bigr(}
\def\){\right)}		\def\BR{\Bigr)}
	\def\BBL{\lb}
	\def\BBR{\rb}
%
%
\def\pt{\etatilde_p}
\def\zb{{\bar{z} }}
\def\zbar{{\bar{z} }}
\def\frac#1#2{{#1 \over #2}}
\def\inv#1{{1 \over #1}}
\def\half{{1 \over 2}}
\def\d{\partial}
\def\der#1{{\partial \over \partial #1}}
\def\dd#1#2{{\partial #1 \over \partial #2}}
\def\vev#1{\langle #1 \rangle}
\def\ket#1{ | #1 \rangle}
\def\rvac{\hbox{$\vert 0\rangle$}}
\def\lvac{\hbox{$\langle 0 \vert $}}
\def\2pi{\hbox{$2\pi i$}}
\def\e#1{{\rm e}^{^{\textstyle #1}}}
\def\grad#1{\,\nabla\!_{{#1}}\,}
\def\dsl{\raise.15ex\hbox{/}\kern-.57em\partial}
\def\Dsl{\,\raise.15ex\hbox{/}\mkern-.13.5mu D}
%
%
\def\ga{\gamma}		\def\Ga{\Gamma}
\def\be{\beta}
\def\al{\alpha}
\def\ep{\epsilon}
\def\vep{\varepsilon}
\def\la{\lambda}	\def\La{\Lambda}
\def\de{\delta}		\def\De{\Delta}
\def\om{\omega}		\def\Om{\Omega}
\def\sig{\sigma}	\def\Sig{\Sigma}
\def\vphi{\varphi}
%
%
\def\CA{{\cal A}}	\def\CB{{\cal B}}	\def\CC{{\cal C}}
\def\CD{{\cal D}}	\def\CE{{\cal E}}	\def\CF{{\cal F}}
\def\CG{{\cal G}}	\def\CH{{\cal H}}	\def\CI{{\cal J}}
\def\CJ{{\cal J}}	\def\CK{{\cal K}}	\def\CL{{\cal L}}
\def\CM{{\cal M}}	\def\CN{{\cal N}}	\def\CO{{\cal O}}
\def\CP{{\cal P}}	\def\CQ{{\cal Q}}	\def\CR{{\cal R}}
\def\CS{{\cal S}}	\def\CT{{\cal T}}	\def\CU{{\cal U}}
\def\CV{{\cal V}}	\def\CW{{\cal W}}	\def\CX{{\cal X}}
\def\CY{{\cal Y}}	\def\CZ{{\cal Z}}
\def\rvac{\hbox{$\vert 0\rangle$}}
\def\lvac{\hbox{$\langle 0 \vert $}}
\def\comm#1#2{ \BBL\ #1\ ,\ #2 \BBR }
\def\2pi{\hbox{$2\pi i$}}
\def\e#1{{\rm e}^{^{\textstyle #1}}}
\def\grad#1{\,\nabla\!_{{#1}}\,}
\def\dsl{\raise.15ex\hbox{/}\kern-.57em\partial}
\def\Dsl{\,\raise.15ex\hbox{/}\mkern-.13.5mu D}
%
%
\font\numbers=cmss12
\font\upright=cmu10 scaled\magstep1
\def\stroke{\vrule height8pt width0.4pt depth-0.1pt}
\def\topfleck{\vrule height8pt width0.5pt depth-5.9pt}
\def\botfleck{\vrule height2pt width0.5pt depth0.1pt}
\def\Zmath{\vcenter{\hbox{\numbers\rlap{\rlap{Z}\kern
0.8pt\topfleck}\kern 2.2pt
                   \rlap Z\kern 6pt\botfleck\kern 1pt}}}
\def\Qmath{\vcenter{\hbox{\upright\rlap{\rlap{Q}\kern
                   3.8pt\stroke}\phantom{Q}}}}
\def\Nmath{\vcenter{\hbox{\upright\rlap{I}\kern 1.7pt N}}}
\def\Cmath{\vcenter{\hbox{\upright\rlap{\rlap{C}\kern
                   3.8pt\stroke}\phantom{C}}}}
\def\Rmath{\vcenter{\hbox{\upright\rlap{I}\kern 1.7pt R}}}
\def\Z{\ifmmode\Zmath\else$\Zmath$\fi}
\def\Q{\ifmmode\Qmath\else$\Qmath$\fi}
\def\N{\ifmmode\Nmath\else$\Nmath$\fi}
\def\C{\ifmmode\Cmath\else$\Cmath$\fi}
\def\R{\ifmmode\Rmath\else$\Rmath$\fi}

\def\barray{\begin{eqnarray}}
\def\earray{\end{eqnarray}}
\def\beq{\begin{equation}}
\def\eeq{\end{equation}}

\def\n{\noindent}

\def\Tr{\rm Tr} 
\def\xvec{{\bf x}}
\def\kvec{{\bf k}}
\def\kvecp{{\bf k'}}
\def\omk{\om{\kvec}} 
\def\dk#1{\frac{d\kvec_{#1}}{(2\pi)^d}}
\def\2pid{(2\pi)^d}
\def\ket#1{|#1 \rangle}
\def\bra#1{\langle #1 |}
\def\vol{V}
\def\adag{a^\dagger}
\def\rme{{\rm e}}
\def\Im{{\rm Im}}
\def\pvec{{\bf p}}
\def\fermiS{\CS_F}
\def\cdag{c^\dagger}
\def\adag{a^\dagger}
\def\bdag{b^\dagger}
\def\vvec{{\bf v}}
\def\muhat{{\hat{\mu}}}
\def\vac{|0\rangle}
\def\pcut{{\Lambda_c}}
\def\chidot{\dot{\chi}}
\def\gradvec{\vec{\nabla}}
\def\psitilde{\tilde{\Psi}}
\def\psibar{\bar{\psi}}
\def\psidag{\psi^\dagger} 
\def\m{m_*}
\def\up{\uparrow}
\def\down{\downarrow}
\def\Qo{Q^{0}}
\def\vbar{\bar{v}}
\def\ubar{\bar{u}}
\def\smallhalf{{\textstyle \inv{2}}}
\def\smallsqrt{{\textstyle \inv{\sqrt{2}}}}
\def\rvec{{\bf r}}
\def\avec{{\bf a}}
\def\pivec{{\vec{\pi}}}
\def\svec{\vec{s}} 
\def\phivec{\vec{\phi}}
\def\daggerc{{\dagger_c}}
\def\Gfour{G^{(4)}}
\def\dim#1{\lbrack\!\lbrack #1 \rbrack\! \rbrack }
\def\qhat{{\hat{q}}}
\def\ghat{{\hat{g}}}
\def\nvec{{\vec{n}}}
\def\bull{$\bullet$}
\def\ghato{{\hat{g}_0}}
\def\r{r}
\def\deltaq{\delta_q}
\def\gcharge{g_q}
\def\gspin{g_s}
\def\deltas{\delta_s}
\def\gQC{g_{AF}} 
\def\ghatqc{\ghat_{AF}}
\def\xqc{x_{AF}}
\def\mhat{\hat{m}}
\def\xup{x_2}
\def\xdown{x_1}
\def\sigmavec{\vec{\sigma}}
\def\xopt{x_{\rm opt}}
\def\Lambdac{{\Lambda_c}}
\def\angstrom{{{\scriptstyle \circ} \atop A}     }
\def\AA{\leavevmode\setbox0=\hbox{h}\dimen0=\ht0 \advance\dimen0 by-1ex\rlap{
\raise.67\dimen0\hbox{\char'27}}A}
\def\ratio{\gamma}
\def\Phivec{{\vec{\Phi}}}
\def\singlet{\chi^- \chi^+} 
\def\mhat{{\hat{m}}}

\def\Im{{\rm Im}}
\def\Re{{\rm Re}}

\def\xstar{x_*}
\def\sech{{\rm sech}}
\def\Li{{\rm Li}}
\def\dim#1{{\rm dim}[#1]}
\def\ep{\epsilon}
\def\free{\CF}
\def\Fhat{\digamma}
\def\ftilde{\tilde{f}}
\def\muphys{\mu_{\rm phys}}
\def\xitilde{\tilde{\xi}}
\def\CI{\mathcal{I}}
\def\nhat{\hat{n}}
\def\ef{\epsilon_F}
\def\as{a_s}
\def\diffk{|\kvec - \kvec' |}
\def\bfT{{\bf T}}

\def\bfC{{\bf C}}
\def\bfP{{\bf P}}

\def\Ct{ \tilde{C} }
\def\Tt{ \tilde{T} }
\def\Pt{ \tilde{P} }
\def\etatilde{\tilde{\eta}}
\def\pt{\tilde{p}}
\def\Ttilde{\Tt}
\def\Ctilde{\Ct}
\def\Ptilde{\Pt}

\def\zbar{{\overline{z}}}
\def\dbar{{\overline{d}}}
\def\pbar{{\overline{\d }}}
\def\Abar{{\overline{A}}}

\def\Z{\mathbb{Z}}

\def\none{\emptyset}

\def\Ctilde{\tilde{C}}
\def\one{{\bf 1}}
\def\utilde{\tilde{u}}


\title{Edge states for  topological insulators in two dimensions  and their Luttinger-like liquids}
\author{Denis Bernard}
\affiliation{Laboratoire de physique th\'eorique, Ecole Normale Sup\'erieure,  CNRS, Paris, France}
\author{Eun-Ah Kim}
\affiliation{Cornell University, Ithaca, NY}
\author{Andr\'e LeClair}
\affiliation{Cornell University, Ithaca, NY}
\affiliation{Laboratoire de physique th\'eorique, Ecole Normale Sup\'erieure,  CNRS, Paris, France}
\affiliation{CBPF,  Rio de Janeiro, Brazil}


\begin{abstract}
Topological insulators in three  spatial dimensions are known to possess a precise
bulk/boundary correspondence,  in that there is a one-to-one correspondence 
between the 5 classes characterized by
bulk topological invariants  and  Dirac hamiltonians on the boundary with 
symmetry protected zero modes.   This holographic characterization of topological 
insulators  is studied in two  dimensions.  
Dirac hamiltonians on the one  dimensional edge   are classified according to the 
discrete symmetries of 
time-reversal,  particle-hole,  and chirality,  extending 
a previous classification in two  dimensions.   We find 17 inequivalent classes,  of 
which 11 have protected zero modes.      Although bulk topological invariants are thus far  known for
only 5 of these classes,  we conjecture that the additional 6 describe edge states of new classes 
of topological insulators.  
 The effects of interactions in two dimensions are  also studied.   We show that all interactions that preserve the symmetry are exactly
marginal,  i.e. preserve the gaplessness.     This leads to a description of the  distinct variations 
of Luttinger liquids that can be realized on the edge.

\end{abstract}

\maketitle

\section{Introduction}
Topological insulators are characterized by bulk wave-functions in $d$ spatial dimensions 
with special topological properties characterized by certain topological invariants,  such 
as the Chern number\cite{Halperin,Chern,Stone,Kane2005,Bernevig15122006,PhysRevLett.98.106803,PhysRevB.76.045302,PhysRevB.79.195321}.
These physical systems possess a kind of holography,  or
bulk/boundary correspondence,  in that they necessarily have protected gapless excitations
on the $\dbar = d-1$ dimensional surface.    These surface modes are typically described by
Dirac hamiltonians.   For example in the integer quantum Hall effect (QHE)  in $d=2$,  the Chern number  is
the same integer as  in the quantized Hall conductivity, and the edge states are chiral Dirac fermions.  

\textcite{Schnyder2008}, \textcite{Ryu2010} and \textcite{kitaev-periodic} classified topological insulators in any dimension 
according to the discrete symmetries  of time reversal $\bfT$,  particle-hole symmetry $\bfC$ and 
chirality $\bfP$ and  found 5 classes of topological insulators in any dimension.     
See also \cite{Wen2}.   These classifications
relied on generic properties in any dimension,  namely the homotopy groups of replica sigma models 
for Anderson localization\cite{Schnyder2008,Ryu2010},  or  the  8-fold periodicity property of spinor representations of $so(n)$  based 
on their Clifford algebras,  which is a mild form of Bott-periodicity in  K-theory\cite{kitaev-periodic}. 
 
The bulk/boundary correspondence was described explicitly in \cite{Schnyder2008}  for $d=3$ spatial dimensions:   using the classification 
of  $\dbar=2$ dimensional Dirac hamiltonians in \cite{Bernard2002},  it was found that precisely 5 of the 13  Dirac classes
had protected surface states with the predicted  discrete symmetries.    In that analysis,  it was crucial that
the classification in \cite{Bernard2002} contained  3 additional classes beyond the 10 Altland-Zirnbauer (AZ)
 classes\cite{AZ},
since it was precisely these additional classes that corresponded to some of the topological insulators.     The reason that 
there are more classes of Dirac hamiltonians is that  AZ classify finite dimensional hermitian matrices (hamiltonians) 
without assuming any Dirac structure.    

In this paper we explore this  `holographic classification'  of topological insulators (TI's)  and topological superconductors (TS's)  in  $d=2$ spatial dimensions,  in order to ascertain whether
it works out as nicely as for $d=3$. 
 The general $d$ dimensional case will be presented elsewhere\cite{Bernard-LeClair2012}.
 It is not obvious from the beginning that this holographic approach 
should reproduce precisely the classifications based on topological invariants.   For instance,  Anderson
localization properties are generally different in $d<2$ verses $d>2$.   
 Also,   we assume that
the surface states can be realized as Dirac fermions,   which is an additional constraint on top of  the discrete symmetries under  consideration. 
More importantly,
 there is no guarantee that there exists a microscopic 
2D model with topological wave-functions with the edge modes we classify.
However  the subsequent holographic classification by two of us\cite{Bernard-LeClair2012}  in arbitrary dimensions 
strengthens the case for the holographic approach as it was found 
using 
only generic properties of Clifford
algebras that this approach gives precisely the known TI's and nothing more in 
odd dimensions. In even dimensions with $d\neq2$, only one additional class with protected surface Dirac fermions were found. The 
 $d=2$ case turned out to be special and it is the focus of this paper.
Also,   it is important to examine this  holographic classification since the edge states are the most experimentally accessible properties.



 This  study requires a classification of Dirac hamiltonians in $\dbar =1$,
which is carried out for the first time  below.  
  We  identify  17  unitarily-inequivalent classes. 
 Since the classifications in \cite{Schnyder2008,Ryu2010,kitaev-periodic} were based on {\it generic} properties in any dimension, 
 it is possible that there exist more classes of topological insulators in $d=2$ due to this richer structure specific to
 $\dbar =1$.    Indeed,  based on our classification,   we find 11 classes of Dirac hamiltonians with protected zero modes
 on the 1 dimensional edge.  
 In addition to the previously  predicted   topological insulators  in classes  A,  C, D,  DIII,   and AII,  
  we find  the  classes AIII,  BDI,  two versions of CII,  
 an additional version of DIII,  and a $\Z_2$ version of D  (the  definition of these classes will be reviewed below; the notation goes back to Cartan).  
 One  interpretation is that, unlike in $d=3$,  for $d=2$  there are classes of $\dbar =1$ Dirac hamiltonians that are protected 
 for reasons other than the existence of a  topological invariant for the $d=2$ band structure.    
 On the other hand,  our new classes could in principle  be characterized by some as yet unknown bulk topological
 invariants.  
 Although this distinction needs to  be kept in mind,  
 henceforth,  for simplicity,  we will refer to all classes with protected zero modes on
 the boundary  as TI's.

   For the QHE,   bulk interactions lead to the fractional 
 QHE,  and the effect of these interactions is that the edge states become Luttinger liquids\cite{Wen}.    
 This is unique to $d=2$ since only in this dimension are quartic interactions on the boundary marginal, which is not
 unrelated to the fact that anyons only exist in 2 dimensions.    
 Thus a criterion for the possible effects of  bulk interactions is the existence of {\it exactly} marginal perturbations 
 of the free  boundary Dirac hamiltonian that are consistent with the discrete symmetries,   since an exactly 
 marginal perturbation deforms the theory but keeps it gapless.    This leads us to also classify quartic,  exactly
 marginal perturbations that are consistent with the discrete symmetries.    
 In addition to the ordinary,  chiral and helical Luttinger liquids,  we find the possibility of 3   additional varieties 
 in the classes DIII and CII.   

The sections below cover the following.   In section II we review the definitions of the 10  AZ classes.  
Section III reviews  the holographic classification of TI  in  $d=3$.    
One-dimensional Dirac hamiltonians are classified in section IV.  This 
classification is completely general,  and could have applications in other areas,  such as
disordered systems.   In section V,  we identify the Dirac theories with protected zero modes,  and
section VI   describes their consequent Luttinger liquids.   

\section{Discrete symmetries}
The  10 Altland-Zirnbauer  (AZ) classes of random hamiltonians   arise when one considers 
time reversal symmetry $(\bfT)$,  particle-hole symmetry $(\bfC)$, and parity or chirality $(\bfP)$.   
These discrete symmetries are defined  to act as follows on a first-quantized hamiltonian $\CH$:
\barray
\nonumber 
\bfT:  ~~~~~~~~~~T \CH^* T^\dagger &=& \CH
\\ 
\bfC: ~~~~~~~~~C \CH^T C^\dagger &=& - \CH 
\label{syms} 
\\ 
\bfP: ~~~~~~~ ~~ ~~P \CH P^\dagger &=& - \CH \nonumber 
\earray
with $T T^\dagger = C C^\dagger = P P^\dagger ={\bf 1} $.    We consider  two hamiltonians $\CH, \CH'$ related by a unitary transformation $\CH' = U \CH U^\dagger$ to be in the same class,  since they have the same eigenvalues.  For $C$ and $T$,  this translates to $C \to C' = UCU^T$  and $T \to T' = UTU^T$.    For $P$,  it amounts to $P \to P'=U P U^\dagger$.  It is thus important to 
identify s these unitary equivalences  in 
order not to over-count classes.   We will sometimes refer to these unitary transformations as gauge transformations.   

For hermitian hamiltonians, $\CH^T = \CH^*$,  thus, up to a sign, $\bfC$ and $\bfT$ symmetries are the same. We focus then on these symmetries
involving the transpose:   $T \CH^T  T^\dagger =  \CH$ and $C \CH^T C^\dagger = - \CH$.   Taking the transpose of this relation, one finds there are two consistent possibilities: $T^T = \pm T,  C^T = \pm C$,  which are unitarily-invariant relations.
  It turns out that unitary transformations allow us to choose $T,C$ to be real;   unitarity of $T,C$ then  implies $C^2 = \pm 1, T^2 = \pm 1$.  The various classes are  thus distinguished by $T^2 = \pm 1, \emptyset$ and $C^2 = \pm 1, \emptyset$,    where $\emptyset$ indicates that the hamiltonian does not have the symmetry,
and  the sign is equivalent 
to the sign in the  relation between $T,C$ and their transpose.  
 One obtains $9 = 3 \times 3$ classes just by considering the 3 cases for $\bfT$ and $\bfC$.    If the hamiltonian has both $\bfT$ and $\bfC$ symmetry, then it automatically has a $\bfP$ symmetry, with $P=TC^\dagger$ up to a phase.  If there is neither $\bfT$ nor
  $\bfC$  symmetry,  then there are two choices $P=\emptyset,1$,  and this gives the  additional class AIII,  leading to a total of 10.   Their properties are shown in Table \ref{tab:AZ}. We also mention that one normally requires $P^2 =1$.  Below, we will require $\bfT$ and $\bfC$ to commute,  thus $P^2 = T^2 { C^\dagger}^2 = \pm 1$.  However one has the freedom $P \to i P$ to restore $P^2 =1$.    In the sequel,  in the cases with both $\bfT, \bfC $ symmetry,  we simply define $P = TC^\dagger$, up to a phase.  

\begin{table}
\begin{center}
\begin{tabular}{|c|c|c|c|}
\hline\hline
AZ-classes &  ~ ~~$T^2$~~ ~ &~~~ $C^2$~ ~~&~~~ $P^2$ ~ ~~ \\
\hline\hline 
A  & $\none$  & $\none$ & $\none$ \\
AIII   & $\none$ & $\none$ & 1  \\
AII  &$ -1 $& $\none$ &$\none$ \\
AI & +1 &$\none$&$\none$ \\
C & $\none$ & $-1$&$\none$ \\
D  & $\none$ & +1 & $\none$  \\
BDI   &+1 & +1 & 1  \\
DIII   & $-1$ & +1 & 1  \\
CII &$ -1$ &$ -1$ & 1 \\
CI & +1 &$ -1$ & 1 \\
\hline\hline 
\end{tabular}
\end{center}
\caption{The 10 Altland-Zirnbauer (AZ) hamiltonian classes. $
\none$ denotes the absence of respective symmetry.  
}
\label{tab:AZ}
\end{table}

\section{Review of the $\dbar =2$ dimensional case}  

The  connection between the bulk topological properties and the existence of protected zero modes
on the boundary was  first pointed out  for  $d=3$ by Schnyder et. al.\cite{Schnyder2008}.   This relied on 
the classification of $\dbar =2$ dimensional Dirac hamiltonians found by two of us\cite{Bernard2002}. 
In this section we review this  holographic classification of  $d=3$ TI's since this illustrates what
we are attempting to accomplish in $d=2$.    

If one requires a Dirac structure of the hamiltonian,  
 then the  AZ  classification can be more refined. 
 The most general hamiltonian in $\dbar =2$ dimensions is of the form:
 \beq
 \label{hamd3}
 \CH =  \(
 \begin{matrix}
 V_+ + V_-  & -i \d_\zbar + A_\zbar \cr
 -i \d_z + A_z &  V_+ - V_- \cr
 \end{matrix}
 \)
 \eeq
 where $\d_z = \d_x - i \d_y, \d_\zbar = \d_x + i \d_y$ with $x,y$ the spatial coordinates and $V_\pm, A_{z, \zbar}$  are
 matrices.  
 The above $\CH$ is just a relabeling of   $\CH =  -i \sigma_x \d_x -i \sigma_y \d_y   +  \vec{\sigma} \cdot \vec{V} + V_0$, 
i.e.  the block structure comes from the Pauli matrices $\sigma$.

One then finds the most general form of the $T,C,P$ matrices that preserve the Dirac structure.  
  Thirteen inequivalent  classes were found\cite{Bernard2002}.   In particular,  there exist two inequivalent 
  versions of the chiral  classes  AIII, DIII, and CI,  simply because the discrete symmetries
  can take different forms.  
   In was shown
 in \cite{Schnyder2008}  that precisely 5 of the 13 classes corresponded to the surface states of TI's,  with discrete symmetries
 consistent with the predictions from bulk topology.    
 As argued there,  the criterion for a TI is that $V_-$ has a zero mode,  i.e. $\det \, V_- =0$.  
 This led to the following identification of TI's,  where the nomenclature of \cite{Bernard2002} is given
 in parentheses.  
 As far as the bulk properties,  the are two types of topological invariants,  $\Z$ and $\Z_2$,  which are also indicated.  
 In the holographic approach,  $\Z$ verses $\Z_2$ corresponds to the two ways of obtaining a zero mode,  namely
 $V_-=0$ or $\det V_- = - \det V_-$ for $V_- $ odd dimensional, and the exceptional case CII,  which is also $\Z_2$.  (See section VA for
 a more detailed discussion of these topological identifications.).  
\begin{itemize}
\item {\bf AIII  (1) ,  DIII (5) ,  CI  (6)  }.  ~~  These are the three classes that are doubled in comparison with AZ.  
 For one of the two in each these classes,  the discrete symmetry forces $V_- =0$.     These  are all of type $\Z$
 or $2\Z$.\footnote{The distinction between $\Z$ and $2\Z$ is explained for the $d=2$ case in section V} 
 
 
\item {\bf AII   ($3_+$)}.  ~~   Here the discrete symmetries require $V_-^T = - V_-$,  which implies that if $V_-$ is odd-dimensional,  $\det \, V_- = 0$.   
 Type $\Z_2$.  
 
 
 \item {\bf  CII  ($9_-$)}. ~~   In this case, the discrete symmetries constrain $V_- =  
 \( 
 \begin{smallmatrix}
 0 & v_- \cr
 w_- & 0 \cr
 \end{smallmatrix}
 \) $   with $v_-^T = - v_- ,  w_-^T =  -w_- $.   Thus if $v_- , w_- $ are odd-dimensional,  then  up to a sign, 
 $\det \, V_- =  \det \, v_- \det \,  w_- =0$.     Type $\Z_2$.

\end{itemize}

\section{The $\dbar =1$ dimensional classification of Dirac hamiltonians}  
\label{sec:1d}
In this section,  we present the complete classification of $\dbar =1$ dimensional Dirac hamiltonians. 
Although the  identification of TI's and TS's will be the subject of the next section,  
it is useful to motivate what follows   
by  discussing   chiral Dirac Hamiltonians with only right moving or left moving fermions\footnote{Note that the  use of  the word ``chiral'' here is different from the terminology of AZ. In  the AZ classification,  classes AIII, BDI, and CII were collectively referred to as chiral classes}. 
Since a mass term necessarily couples left and right movers (see section V),  these classes have a protected zero mode for 
somewhat trivial reasons. 
Such Hamiltonians cannot be realized on a 1d  lattice and they necessarily break {\bf T} and {\bf P}. However they can appear as a $\dbar=1$-edge state of a 2d  TI or TS in classes A, C, and D which break both {\bf T} and {\bf P}.  An example of class A is  the quantum Hall effect. Depending on the number of filled Landau levels there are $\Z$ number of edge states\cite{Halperin}.
 An example of class C is the   spin quantum Hall effect in a singlet time-reversal breaking superconductor. The spin quantum Hall conductivity will be proportional to the Cooper pair angular momentum, hence this is  a $\Z$ TS.  Although there is no known experimental realization, $d_{x^2-y^2}+id_{xy}$ superconductor (SC) was extensively discussed theoretically\cite{Read2000,PhysRevB.60.4245}. A realization of class D would be the  thermal Hall effect of a  time-reversal breaking superfluid of spinless (fully spin polarized) fermions. The  $\nu=5/2$ quantum Hall state could be a $p_x+ip_y$ paired superfluid of composite fermions\cite{Moore:1991kx}.

All non-``chiral'' non-interacting 1d Dirac hamiltonians with equal number of right-movers and left-movers can be written as
$\CH = -i \sigma_x \d_x  + \vec{\sigma} \cdot \vec{A}  +  V_+$, where $\vec{\sigma}$ are the Pauli matrices acting on a space of right/left-movers $|\sigma_x=\pm\rangle$. 
Redefining $A_z = V_-$, these hamiltonians  can be expressed as   
\beq
\label{Hform}
\CH  = \( \begin{matrix}   V_+ +  V_-  & - i \d_x + A \cr- i \d_x + A^\dagger & V_+ - V_- \cr  \end{matrix}  \).
\eeq
The potentials  $V_\pm $ are hermitian matrices and $A=A_x+iA_y$ where  $A_{x,y}$ are also hermitian matrices  in general. The dimension of $V_\pm$ and $A$ is the number of edge mode species for each chirality. When $V_\pm$ and $A$ are even dimensional we use $\vec{\tau}$ to denote a set of Pauli matrices acting on the even dimensional flavor space. $\one$ will denote the identity in either the $\sigma$ or $\tau$ space. Note that $\vec{\sigma}$ and $\vec{\tau}$ have distinct physical meaning: $\vec{\sigma}$ acts on the space of ``chirality'' as we show explicitly in section\ref{subsec:second},  and it is responsible for the block structure of Eq.\eqref{Hform}, whereas  $\vec{\tau}$ acts on the space of flavors which could be spin or pseudo-spin. If there is  spin-momentum locking (see section\ref{subsec:second}) $\vec{\sigma}$ will act on the spin space as well as on the space of ``chirality''.  

The Dirac derivative structure  of $\CH$  constrains the form of  $T,C$, and $P$ in terms of $\vec{\sigma}$ and $\vec{\tau}$.  Furthermore, we can specify the conditions $V_\pm$ and $A$ have to satisfy in order for $\CH$ to have  discrete symmetries under specific $T,C$, or $P$. Hence 
the specific forms of symmetry transformations  can be used to classify hamiltonians of form Eq.\eqref{Hform}. Since, as described
below,   there are multiple sets of matrices $T,C, P$ with the same  $T^2$, $C^2$, $P^2$,  this scheme refines the AZ classification of Table~\ref{tab:AZ}.   Here we find  even  more classes of Dirac hamiltonians in $\dbar=1$ than in  $\dbar =2$,   and more classes with symmetry protected zero modes (see section\ref{sec:TI}).

In the rest of this section, we first specify the forms of  $T,C$ and $P$  symmetry that preserve the Dirac structure,  
 and describe the resulting conditions on $V_\pm$ and $A$ in a  fixed $\vec{\sigma}$ basis and arrive at 25 classes as summarized in Table~\ref{tab:25-17}. We then check for unitary equivalences. The unitary transform is 
\beq
{\cal H} \to U_\theta {\cal H} U_\theta^\dag
\label{eq:unitary}
\eeq
 with $U_\theta $ a rotation about  the $x$-axis in $\sigma$-space
by an angle $\theta$:  
\beq
U_\theta = u\cdot e^{i \theta \sigma_x /2}  = u\cdot(\one\cos( \theta/2)  + i \sigma_x \sin ( \theta/2))
\label{eq:U}
\eeq
 where $u$ is  unitary and  commutes  with $\sigma_x$. We find 17 unitarily-inequivalent classes, each forming a row separated by a horizontal line in Table~\ref{tab:25-17}.

Consider first the $\bfT$ symmetry.     In order to preserve the derivative structure of the hamiltonian Eq.\eqref{Hform},  using 
$(-i\d_x)^T = i \d_x$, one finds that $T$  must  anti-commute with $\sigma_x$.  Since $T$ is (anti)-symmetric and unitary, it is then either proportional to $\sigma_z$ or $i\sigma_y$.
This leads to 2 ways of implementing of $\bfT$-symmetry transformations: using either 
\beq
\label{Ts1}
T =\eta_t\otimes i\sigma_y =\begin{pmatrix}  0  &\eta_t  \cr -\eta_t  &0  \end{pmatrix}
\eeq
\beq
\label{Ts2}
\Tt =  \tilde \eta_t\otimes \sigma_z=
\( \begin{matrix}  \etatilde_t  & 0 \cr  0  & -\etatilde_t \cr  \end{matrix}  \),
\eeq
where $\eta_t$ or $\etatilde_t$ are unitary matrices in general. 
Then, for a hamiltonian of form Eq.\eqref{Hform} to have $\bfT$ symmetry the potentials have to satisfy either 
\beq
\eta_t\, V_\pm^T = \pm  V_\pm\, \eta_t  ,\quad \eta_t\, A^T = - A\, \eta_t 
\eeq
or 
\beq
\etatilde_t\,  V_\pm^T =   V_\pm\,  \etatilde_t  ,  \quad  \etatilde_t\,  A^*  = - A\,  \etatilde_t  
\eeq
Now the condition $T^T = \pm T$ ($T^2=\pm1$) which distinguishes AI from AII for instance, implies either $\eta_{t}^T = \pm \eta_{t}$ or  $\etatilde_{t}^T =\pm \etatilde_t$. Hence all AZ classes with $\bfT$-symmetry are further refined depending on whether $T$ (Eq.\eqref{Ts1}) or $\Tt$ (Eq.\eqref{Ts2}) is used to implement $\bfT$. This distinction has a physical significance:
the use of $T\propto i\sigma_y$ leads to spin-momentum locking (see section~\ref{subsec:second}). 

 \begin{table*}[t]
\begin{tabular}{|c|c|c|c|c|c|c|}
\hline\hline
1d-classes  & ~~~ $\bfT$  ~~~      & ~~~ $\bfC$  ~~~         &   ~~~ $\bfP$ ~~~  &        ~~~ $V_\pm$ ~~~   & ~~~$A$ ~~~ & {\bf zero-mode} \\
\hline\hline 
A                             	&   $\none$                       	&   $\none$                       	& $\none$                      	&$V_\pm^\dag=V_\pm $& 
&  $\Z$  \\
\hline
AIII$_{(1)} $           	&   $\none$                       &   $\none$             &${\bf 1}\otimes\sigma_z$  	& $V_\pm=0$		&
& ${\Z}$    \\
AIII$_{(1)}'$           	&   $\none$                       &   $\none$             & ${\bf 1}\otimes i \sigma_y$  	&  $V_+=0$ 		&
   &  \\
\hline
AIII$_{(2)} $           	&   $\none$                       &   $\none$          & $\tau_z\otimes\sigma_z$   	&$\tau_zV_\pm=-V_\pm\tau_z$&$\tau_z A=A\tau_z$  & \\
AIII$_{(2)}' $           	&   $\none$                       &   $\none$          &  $\tau_z\otimes i \sigma_y$   	&$\tau_zV_\pm= \mp V_\pm\tau_z$&
     &    \\
\hline AII$_{(1)}$  &${\bf 1}\otimes i\sigma_y$   &  $\none$                         	& $\none$                    	&$V_\pm=\pm V_\pm^T$&$A^T=-A$   &$\Z_2$ \\
\hline AII$_{(2)}$  &$i\tau_y\otimes \sigma_z$   &   $\none$                       	& $\none$           &$\tau_yV_\pm^T=V_\pm\tau_y$&$\tau_yA^*=-A  \tau_y$      &   \\
\hline AI$_{(1)} $  &$i\tau_y\otimes i\sigma_y$   &   $\none$                       	& $\none$         &$\tau_yV_\pm^T=\pm V_\pm\tau_y$&$\tau_yA^T=-A\tau_y$     &  \\
\hline AI$_{(2)} $  &${\bf 1}\otimes \sigma_z$   	&   $\none$                      	& $\none$                       	&$V_\pm^T=V_\pm$&$A^*=-A$        & \\
\hline 
C              			& $\none$  		& $i\tau_y\otimes{\bf 1} $ 	& $\none$             		&$\tau_yV^T_\pm=-V_\pm\tau_y$&$\tau_y A^*=-A\tau_y$       &$2\Z$  \\
C$'$                   & $\none$  		& $i\tau_y\otimes \sigma_x $ 	& $\none$             		&$\tau_yV^T_\pm= \mp V_\pm\tau_y$&$\tau_y A^T=-A \tau_y$    &    \\
\hline D                 	   &   $\none$      		 &${\bf 1}\otimes{\bf 1} $  	& $\none$                        	&$V_\pm= -  V_\pm^T$&$A^*=-A$     &   ${\Z , ~\Z_2}$ \\
D$'$                	 &   $\none$      		 &${\bf 1}\otimes \sigma_x $  	& $\none$                        	&$V_\pm =\mp V_\pm^T $&$A^T=-A $      &  \\
\hline BDI$_{(1)}$ &$i\tau_y\otimes i\sigma_y$ &${\bf 1}\otimes{\bf 1}$ &$ i\tau_y \otimes i \sigma_y $     &$V_\pm =-V_\pm^T =\mp \tau_yV_\pm \tau_y$&$A= -A^* = -\tau_y A^T \tau_y$   &\\
BDI$_{(1)}'$ &$ i \tau_y \otimes i\sigma_y$ &$\tau_x \otimes \sigma_x$ &$ \tau_z \otimes \sigma_z $  &$  V_\pm = \pm \tau_y V_\pm^T \tau_y   = \mp \tau_x V_\pm^T \tau_x $&$\tau_{x,y} A^T = - A  \tau_{x,y}$      & \\

\hline BDI$_{(2)}$ &${\bf 1}\otimes \sigma_z$ &${\bf 1}\otimes{\bf 1}$ & $ 1 \otimes \sigma_z $    	&$V_\pm=0$&$A^*=-A$      & ${\Z}$ \\

\hline DIII$_{(1)}$ & ${\bf 1}\otimes i\sigma_y$  &${\bf 1}\otimes{\bf 1}$  &$ 1 \otimes i \sigma_y $     &$V_+=0,\, V_-^T=-V_-$	& $A=-A^*=-A^T$ & ${\Z_2}$ \\
\hline DIII$_{(2)}$  &$i\tau_y\otimes \sigma_z$   &${\bf 1}\otimes{\bf 1}$  &$ i\tau_y \otimes \sigma_z  $  &$ V_\pm = - V_\pm^T = - \tau_y V_\pm \tau_y $&$A= -A^*  = - \tau_yA^T \tau_y $   &  ${\Z_2}$ \\
DIII$_{(2)}'$ & $i \tau_y \otimes \sigma_z$  & $\tau_x \otimes \sigma_x$  & $\tau_z \otimes i \sigma_y $   &$ V_\pm = \tau_y V_\pm^T \tau_y = \mp \tau_x V_\pm^T \tau_x$	& $ A= - \tau_y A^* \tau^y = - \tau_x A^T \tau_x $  &\\

\hline 
CII$_{(1)}$  &${\bf 1}\otimes i\sigma_y$  &$i\tau_y\otimes{\bf 1}$  & $i\tau_y \otimes i \sigma_y$       &$V_\pm=\pm V_\pm^T = \mp \tau_yV_\pm \tau_y$&$A=-A^T= - \tau_yA^* \tau_y$ & ${\Z_2}$\\
CII$_{(1)}'$  &$    \tau_x \otimes i \sigma_y      $  &$  i\tau_y \otimes \sigma_x          $ & $\tau_z \otimes \sigma_z$  &$ V_\pm = \pm \tau_x V_\pm^T \tau_x = \mp \tau_y V_\pm^T \tau_y  $   &$ \tau_{x,y} A^T = - A \tau_{x,y}   $ & \\

\hline CII$_{(2)}$  &$i\tau_y\otimes \sigma_z$  &$i\tau_y\otimes{\bf 1}$ &$ 1 \otimes \sigma_z$              	&$V_\pm=0$&$A =-\tau_yA^* \tau_y$  & ${2\Z}$\\
\hline CI$_{(1)}$ &$i\tau_y\otimes i\sigma_y$ &$i\tau_y\otimes{\bf 1}$ & $1 \otimes i \sigma_y$&$V_+=0,\, \tau_yV_-^T=-V_-\tau_y$&$A=-\tau_y A^T \tau_y= - \tau_y A^* \tau_y$   & \\
\hline CI$_{(2)}$  &${\bf 1}\otimes \sigma_z$ &$i\tau_y\otimes{\bf 1}$	& $i\tau_y \otimes \sigma_z$     &$V_\pm=V^T_\pm=-\tau_yV_\pm\tau_y$&$ A=-A^* = -\tau_y A^* \tau_y  $   & \\
CI$_{(2)}'$  &$    \tau_x \otimes \sigma_z     $  &$ i\tau_y \otimes \sigma_x   $ & $ \tau_z \otimes i \sigma_y$   &$   V_\pm = \tau_x V_\pm^T \tau_x = \mp \tau_y V_\pm^T \tau_y  $ &$ A= -\tau_x A^* \tau_x = - \tau_y A^T \tau_y  $ &  \\
\hline\hline 
\end{tabular}
\caption{The properties of the 25 non-chiral $\dbar=1$ Dirac classes. 17 unitarily-inequivalent classes separated from each other by a horizontal line. The first column lists the $\bar d=1$ Dirac classes.
Columns $\bfT$, $\bfC$ and $\bfP$ show representations of symmetry transformations for each class.  The columns  $V_\pm$ and $A$  show symmetry constraints on the potentials. A blank cell denotes absence thereof. The symmetry constraints guarantee zero modes in some classes (see section~\ref{sec:TI}). The last column shows classes with symmetry protected zero modes and the type of zero modes.   }
\label{tab:25-17}
\end{table*}

  Finally we can choose representations of $\eta_t$ in terms of $\vec{\tau}$ up to the unitary transformations: $\eta_t = {\bf 1}$ 
if $\eta_t^T = \eta_t$, and $\eta_t = i\tau_y$ if $\eta_t^T = - \eta_t$\cite{WignerDyson}. We can do the same for $\etatilde_t$. The unitary transform $T \to U T U^T$  corresponds to $\eta \to u \eta u^T$  with $u$ unitary,  for all $\eta$'s. The 
unitary transformation affects the choice of  $\one$ v.s. $\tau_x$ for $\eta_t$'s.
However the unitary transform cannot affect the distinction between $T$ and $\Tt$. In particular
when $\bfT$ is the only available discrete symmetry, $T^2, \Tt^2=\pm1$ completely classifies $\dbar=1$ Dirac Hamiltonians into AI$_{(1)}$, AI$_{(2)}$ and AII$_{(1)}$, AII$_{(2)}$. See Table~\ref{tab:25-17}. 

We can specify $C$, following steps analogous to those for specifying $T$. As $C$ must commute with $\sigma_x$ for Dirac hamiltonian Eq.\eqref{Hform},  it is in the linear span of  $\one$ and $\sigma_x$. Hence
there are two possibilities:
  \begin{align}
\label{Cs1}
 & C = \eta_c\otimes\sigma_x, 
\quad \eta_c\,  V_\pm^T \!=\! \mp V_\pm \, \eta_c,\; \eta_c \,  A^T\! =\! - A\,  \eta_c\\
& \Ct  = \tilde\eta_c\otimes \one, 
\quad\etatilde_c\,  V_\pm^T \!=\! - V_\pm\,  \etatilde_c, \; \etatilde_c \,  A^* \!=\! - A\,  \etatilde_c
\nonumber
\end{align}
with $\eta_c$ and $\etatilde_c$ unitary.    
The condition $C^T = \pm C$ that distinguishes AZ class C from D for instance, implies that $\eta_{c}^T = \pm \eta_{c}$  or  $\etatilde_c^T=\pm \etatilde_c$. One can again represent up to unitary transformations $\eta_c = {\bf 1}$ if $\eta_c^T = \eta_c$, and 
$\eta_c = i\tau_y$ if $\eta_c^T = - \eta_c$. This again refines the  AZ classes with $\bfC$ symmetry. However unlike $T$ and $\Tt$ which are unitarily-inequivalent, $C$ and $\Ct$ are unitarily-equivalent for non-zero $A_y$ (see the end of this section). We denote such unitarily-equivalent refinements using primed notation within the same row in  Table~\ref{tab:25-17}. In particular, this completes our classification of $\dbar=1$ Dirac hamiltonians with only $\bfC$ symmetry into C, C', D, D'.

Consider now $\bfP$ symmetry.   $P$ must anti-commute with $\sigma_x$ for the  Dirac hamiltonian Eq.\eqref{Hform}, so $P$ is in the linear span of $\sigma_y$ and $\sigma_z$. For $P$ unitary, this implies that $P=\eta_p\cdot(\cos b\ \sigma_y +\sin b\ \sigma_z)$ for some real $b$. All these choices  are unitarily-equivalent by rotations around the $x$-axis in the sigma space. However, 
  in order   to accommodate $ P = T C^\dagger$  in all cases, we define two unitarily-equivalent types:
\begin{align}
\label{Ps}
P  &=  \eta_p\otimes \sigma_z
\quad \eta_p\,  V_\pm =-   V_\pm\,  \eta_p, \;  \eta_p\,   A  =  A\,  \eta_p 
\\  \nonumber
\Pt   &=   \etatilde_p\otimes  i  \sigma_y
\quad\etatilde_p\,  V_\pm = \mp  V_\pm\,  \etatilde_p, \; \etatilde_p\,   A^\dagger   =  A\,  \etatilde_p 
\end{align}
where $\eta_p$ and $\etatilde_p$ are unitary. 
The unitary freedom reduces to  $\eta_p \to u \eta_p u^\dagger$ and the same for $\etatilde$.
Up to unitary transformations there are two choices: $\eta_p,  \etatilde_p = 1$ or $\tau_z$.   This gives 4  AIII classes.

Finally for the classes with both $\bfT, \bfC$ symmetries,   $\bfT$ and $\bfC$  must either commute
or anti-commute\cite{Wen2}.   The argument is simple.   Given both $T$ and $C$,  a $P$ symmetry is provided by
$P= T C^\dagger$  or  $P=C^\dagger T$.    These two $P$'s must be equivalent up to a sign since $P^2=1$,  
thus $T C^\dagger = \pm C^\dagger T$,  which is a gauge-invariant condition.
   Thus $T,C$  commute or anti-commute,  since in all cases,  $C^\dagger = \pm C$.

Now the  AZ classes BDI, CI, DIII, and CII refines into 12 classes; among these 8 are gauge inequivalent.
We label the  three subclasses associated with the  BDI class by BDI$_{(1)}$, BDI$_{(2)}$, BDI$_{(2)}'$,
and similarly for  CI, DIII, and CII. 
Table~\ref{tab:25-17} shows this classification with respective representations of $\bfT$, $\bfC$ and $\bfP$. In some cases   $\eta_t $ or $\eta_c$ had to be taken to be $\tau_x$ which is unitarily-equivalent to $\one$, 
in order for $\bfT$ and $\bfC$  to anti-commute. 
When there are both $\bfT, \bfC$ symmetries, then there is automatically a $P = TC^\dagger$ symmetry (up to a phase).
 Depending on the type of $C,T$,  one finds the $\Z_2$ graded multiplication:  
$P = T C^\dagger,  P = \Tt \Ct^\dagger,  \Pt = T \Ct^\dagger,  \Pt = \Tt C^\dagger $.  
This gives $\eta_p=\eta_t \eta_c^\dagger$ or  $\etatilde_t \etatilde_c^\dagger$ and  $\etatilde_p=  \eta_t \etatilde_c^\dagger$ or  
$\etatilde_t \eta_c^\dagger$.  

 \begin{table*}
\begin{tabular}{|c|c|c|c|}
\hline
\hline
$\dbar=1$ classes & zero modes&topological invariant & examples\\
\hline
\hline
A & $\Z$ & $\Z$ & QH edge states \\
C & $2\Z$ &$2\Z$  & spin QH edge states  in $d+id$-wave SC\cite{Read2000,PhysRevB.60.4245}\\
D & $\Z$ & $\Z$ & thermal QH edge states in spinless chiral $p$-wave SC\cite{Read2000}\\
\hline\hline
\end{tabular}
\caption{$\dbar=1$ chiral Dirac hamiltonian classes.  \label{tab:chiral}
}
\end{table*}

\begin{table*}
\begin{tabular}{|c|c|c|c|c|c|c|c|}
\hline\hline
$\dbar=1$ classes &\bfT&\bfC&\bfP& zero modes & top.  inv. & locking & examples\\
\hline\hline
{\bf \color{red}AIII$_{(1)}$} & $\none$&$\none$&$\sigma_z$ &$\Z$ & &  &   \\
\hline
AII$_{(1)}$ &$i\sigma_y$ &$\none$&$\none$& $\Z_2$ &$\Z_2$ &Y & HgTe/(Hg,Cd)Te\\
\hline
{\bf \color{red}D  }& $\none$&$\one$&$\none$&$\Z_2$  && & \\
\hline
{\bf \color{red} BDI$_{(2)}$ }&$\sigma_z$&$\one$&$\sigma_z$ &$\Z$ &&&   \\
\hline
DIII$_{(1)}$ & $i\sigma_y$&$\one$&$i\sigma_y$&$\Z_2$ &$\Z_2$& Y&$(p+ip)\times(p-ip)$-wave SC\\
\hline
{\bf \color{red} DIII$_{(2)}$} & $i\tau_y\otimes\sigma_z$&$\one$&$i\tau_y\otimes\sigma_z$&$\Z_2$ &$\Z_2$& N &particle-hole symmetric KM model\\
\hline
{\bf \color{red} CII$_{(1)}$} &$\one\otimes i\sigma_y$& $i\tau_y\otimes\one$&$i\tau_y\otimes i\sigma_y$&$\Z_2$ && Y& doubled KM  \\
\hline
{\bf \color{red} CII$_{(2)}$ }& $i\tau_y\otimes \sigma_z$&$i\tau_y\otimes\one$&$\one\otimes \sigma_z$ &$2\Z$ && N & trigonally strained graphene\cite{Levy30072010}\\
\hline\hline  
\end{tabular}
\caption{$\dbar=1$ non-chiral Dirac hamiltonian classes with symmetry protected zero modes. The spin-momentum locking column is left blank when spins cannot be assigned because the time-reversal operator do not involve either $i \sigma_y$ or $i\tau_y$.
New classes are shown in boldface (red online).  
\label{tab:non-chiral}}
\end{table*}

Let us finally return to the issue of unitary equivalence.  The unitary transform of Eq.~\eqref{eq:unitary} preserves the Dirac structure for $U_\theta$ of Eq.~\eqref{eq:U}. The  two possibilities $T$ and $\Tt$ for $\bfT$ are unitarily-inequivalent, because unitary transformations preserve the relation $T^T=\pm T$,  or equivalently,   $U_\theta\sigma_{y,z} U_\theta^T=\sigma_{y,z}$. However $C$ and $\Ct$ are unitarily-equivalent for non-zero $A_y$,  since 
$U_{\pi/2}  \sigma_x  U_{\pi/2}^T = i$. In  Table~\ref{tab:25-17}, we listed all 25 classes separating 17 unitarily-inequivalent classes by horizontal lines. It is important to note however that all of the 25 classes should be viewed as inequivalent once $U_\theta$ is used to set $A_y=0$ since $C, \Ct$ are inequivalent under the residual symmetry. (If $A_y=0$, $A^* = A^T$.) We will take this route in the next section where we investigate the symmetry protection of zero modes.

\section{``Topological Insulators''  in two dimensions}
\label{sec:TI}
We conjecture a `holographic' classification of 2D TI-TS based on  the classification of $\dbar=1$ Dirac hamiltonians that are symmetry protected to be gapless,  i.e. have a protected zero mode.  We list such $\dbar=1$ Dirac hamiltonian classes in Tables \ref{tab:chiral} and \ref{tab:non-chiral}.
For a {\it subset} of these classes, there exists  a $d=2$ gapped hamiltonian in the same class and a known topological invariant 
which one can calculate from the  ground state wave function which takes on $\Z$-values or $\Z_2$-values\cite{Schnyder2008, Ryu2010};  these are indicated in the columns denoted ``topological invariant''.    Surprisingly, for a class with a known bulk topological invariant, there is a correspondence between the values it can take and the number of gapless Dirac edge branches (dimension of  the block matrices Eq\eqref{Hform} for  the non-chiral case). 
Namely, classes with $\Z$-invariants are gapless for any number of Dirac edge branches;  classes with $\Z_2$-invariants  are gapless only when there are odd-number of branches for each chirality. 
The main point of this paper is that there are additional classes with protected edge zero modes beyond the 5 predicted on the
basis of the known  topological invariants.   

In the rest of this section we enumerate  the classes  of $\dbar =1 $ Dirac hamiltonians  that  have a protected zero mode as a consequence of the
discrete symmtries.
 We then comment on the  microscopic 2d  models  corresponding to a subset of our  new classes. We finally discuss physical properties of these classes such as spin-momentum locking through a second quantized description.

\subsection{First quantized description}
First we discuss the chiral (only right or left moving) Dirac fermion classes we mentioned at the beginning of section~\ref{sec:1d}. These are massless for a ``trivial'' reason since a mass term necessarily couples left to right. As  $\bfT$ and $\bfP$ transform left to right movers (see below),   hamiltonians with these symmetries cannot be chiral. On the other hand, AZ classes A, C, D  have at most a $\bf C$ symmetry and can be chiral. For chiral hamiltonians in classes A, C, D, any $\Z$ number of branches will be gapless.  For chiral class C,  since the auxiliary $\tau$ space is doubled, as explained above this is of type $2\Z$.
See Table~\ref{tab:chiral} for the summary. 

Now consider non-chiral hamiltonians of the  form Eq.~\eqref{Hform} whose  block diagonal structure implies that the second quantized theory has both right movers $\psi_R\equiv\langle x|\sigma_x=+ \rangle$ and left movers $\psi_L\equiv\langle x|\sigma_x=- \rangle$  (see below).  The hamiltonian $\CH$  is gapless if it has 
a zero eigenvalue at $\kvec=0$, 
i.e.  $\det \, \CH (\kvec =0)  = 0$. Below we simplify this into a condition on $V_-$.

The potential $A_x$ can be removed by redefining the fields in the second quantized theory:
$\psi_{L,R}  \to e^{-i \int^x  A_x (x) dx } \psi_{L,R}$ (see subsection~\ref{subsec:second}).    
A  constant $V_+$ is a chemical potential which shifts the overall energy levels. Hence we set this to zero. Now the condition for existence of a zero mode and hence a gapless spectrum is 
\beq
\label{TIdef}
\det   
\(
\begin{matrix}
V_-  &  i A_y \cr 
- i A_y &  - V_-\cr
\end{matrix} \) = 0 
\eeq
However Eq.~\eqref{TIdef} is difficult to use in general~\footnote{Only when $V_- $ and $A_y$ commute does  Eq.~\eqref{TIdef} equal  $\det ( V_-^2 + A_y^2)=0$ which factorizes into $ \det\,( V_- + i A_y) \det\, (V_- -i A_y)=0$. This offers a unitary transform independent criterion for gaplessness in terms of potentials: $\det\, (V_- + i A_y )=0$. }.
Hence we use the freedom of unitary transform $U_\theta$  to set $A_y=0$. 
The criterion for a TI is now simply $\det \,  V_- = 0$ for fixed $A_y=0$. 

Now we test if the conditions on $V_-$ imposed by symmetry listed in Table~\ref{tab:25-17} guarantee $\det\, V_-=0$. As the choice of $A_y=0$ makes $C$ and $\Ct$ inequivalent we consider all 25 entries. Once we identify symmetry protected gapless Dirac classes, we check for unitary equivalence among those by consulting the Table~\ref{tab:25-17}. In Table~\ref{tab:non-chiral} we list unitarily inequivalent protected classes.  


There are two generic  types of constraints on $V_-$  that protect  a gapless spectrum. First, $V_-=0$ guarantees  $\det\, V_-=0$ independent of the dimension of $V_-$ nor  the $\Z$  number of edge modes.   This is identified with a type $\Z$  TI.   If the $\bfT$ or $\bfC$ symmetry involves a doubling
of the auxiliary $\tau$ space,   then this doubling is the signature of a type $2\Z$ TI.\cite{Bernard-LeClair2012} 
 Second, $V_-^T=-V_-$ implies $\det V_-=-\det V_-$ when 
$V_-$ is {\it odd dimensional}, and  hence $\det \, V_- =0$.    
By analogy with the 3d case,   those that rely on $V_-^T = - V_-$
with $V_-$ odd-dimensional should  be  of $\Z_2$ type because of the even/odd aspect. 

There are also two exceptional cases:  
\noindent
\medskip
{\bf DIII$_{(1)}$~~~}   Here 
 $\etatilde_t = i\tau_y$, $\etatilde_c = \one$,  $\eta_p = i\tau_y$.    Here $V_-^T = - V_-$, however it is even dimensional,  
and constrained to be of the form $V_- = \(  \begin{smallmatrix} a_-  & b_- \cr b_- & - a_- \cr \end{smallmatrix} \)$ with 
$a_-^T = - a_- ,   b_-^T = - b_-$.   Thus,  if $a_-, b_-$ are one  dimensional,   then  $V_- =0$.    Type $\Z_2$.

\medskip
\noindent
{\bf CII$_{(1)}$~~~}   Here  $ \eta_t = \tau_x,   \eta_c =  i \tau_y,  \eta_p = - \tau_z$.   
$
V_- = \(
\begin{smallmatrix}
0 & b_- \cr
c_- & 0 \cr
\end{smallmatrix}
\)
$
with $b_-^T = - b_-,  c_-^T = - c_-$.   If $b_-, c_-$ are odd dimensional,  then,  up to a sign,  $\det \, V_- = \det \, b_-  \det c_- =0$.  
Type $\Z_2$.

The table \ref{tab:non-chiral} lists  new classes with protected Dirac edge modes in boldface(red online). An immediate question is whether these classes can be realized in a microscopic 2D model and if so,  why  they were missed in previous classifications. First we point out that by considering an
additional reflection symmetry,  Yao and Ryu\cite{Yao-Ryu}  recently found  topological invariants for all of our new classes except CII$_(1)$.  As first noticed by Fu \cite{Fu2011}, when considering microscopic realizations of topological insulators,  point-group symmetry can play an  important role. While we required our non-chiral edge state to be described by a  Dirac hamiltonian,   it is plausible that  the latter assumption  automatically implies a  reflection symmetry for some of the classes for $\dbar=1$. This is a topic to be  investigated  further in the future. Nevertheless, what is clear from the work\cite{Yao-Ryu} 
 is that indeed there are microscopic 2d  theories whose edge states are described by our new classes.  

 Turning to physical realizations of the new classes of edge states so far we have found two examples: {DIII}$_{(2)}$ and {CII}$_{(2)}$. An example of 
 {DIII}$_{(2)}$ is the Kane-Mele model in the presence of particle hole symmetry\cite{Zheng2011,Hohenadler2011a}. This can be viewed aa special case of {AII}$_{(1)}$-type TI with additional particle-hole symmetry. The additional symmetry enables quantum Montecarlo simulations without sign-problems. But it also means absence of spin or charge edge current as we will discuss  further in the next section. Of particular interest is the  zero field QHE in trigonally strained graphene\cite{Levy30072010,PhysRevB.76.045430} as an example of  {CII}$_{(2)}$. The details of this identification will be presented elsewhere\cite{Hsu-Kim2012}. However,  the underlying reasoning is rather simple. 
The observation of Landau levels 
in \cite{Levy30072010} in the absence of magnetic field calls for a $\Z$ type TI among time-reversal symmetric classes. In the original classification by \textcite{Schnyder2008} $\Z$ type TI are found only among {\bf T} breaking classes. Since trigonal strain introduces pseudo-magnetic fields of opposite direction for two valleys, there are $2\Z$ edge modes when the system is subject to a  confining potential. 

\subsection{Second quantized description and spin-momentum locking}
\label{subsec:second}

One can define a second-quantized hamiltonian:
\beq
\label{2nd}
H =  \int dx  ~ \sum_{a,b}  \psi^\dagger_a (x)  \CH_{ab} \psi_b (x) 
\eeq
from $\CH$ of Eq.~\eqref{Hform}. Now let $\bfT, \bfC$ be time-reversal and particle-hole transformation operators in the
field theory and define 
\beq
\bfT  \psi_a \bfT^\dagger =  T_{ab} \psi_b, ~~~~~
\bfC  \psi_a \bfC^\dagger =  C_{ab}   \psi^\dagger_b.
\eeq
This and the $T,C$ properties of $\CH$ (Eq.~\eqref{syms}) implies the invariance:
 $\bfT H \bfT^\dagger \!=\! H$,  $\bfC H \bfC^\dagger \!=\! H$.

Since right movers and left movers are $\psi_R\equiv\langle x|\sigma_x=+ \rangle$ and left movers $\psi_L\equiv\langle x|\sigma_x=- \rangle$, the spinor field $\psi$ has the 
block structure:
\beq
\label{block} 
\psi =  \(  \begin{matrix}   \psi_R  +  \psi_L  \cr  \psi_R - \psi_L \cr \end{matrix}  \)
\eeq
in the eigenbasis of $\sigma_z$. Upon passing to Euclidean space by $t\to -i\tau$, the Schrodinger equation for $\CH$ in Eq.~\eqref{Hform}, $i \d_t \psi  =  \CH  \psi$,  becomes $\d_z \psi_R =   \d_\zbar \psi_L =0$,  where $\d_\zbar = \d_\tau + i \d_x,   \d_z = \d_\tau - i \d_x $. This confirms the anticipated chirality of $\psi_R$ and $\psi_L$. 

The $\bfT$ and $\bfP$ transformations exchange left and right movers:
\begin{align}
\label{Ttrans} 
T: &\quad  \psi_R \to - \eta_t \psi_L , \quad\psi_L  \to \eta_t \psi_R\nonumber\\
\Ttilde:   &\quad \psi_R \to \etatilde_t\psi_L  ,  \quad\psi_L \to \etatilde_t \psi_R
\end{align}
 and 
\begin{align}
P:   &~~~~ \psi_R \to  \eta_p  \psi_L  ,  ~~~~~\psi_L \to \eta_p  \psi_R
\nonumber\\ 
\Ptilde:    &~~~~ \psi_R \to - \etatilde_p   \psi_L,  ~~~~~\psi_L \to \etatilde_p  \psi_R
\end{align}
On the other hand, $C$ transforms fields into their conjugates:
\begin{align}
\label{Ctrans}
C: &~~~~ \psi_R  \to \eta_c \psi_R^\dagger,  ~~~~~ \psi_L \to - \eta_c  \psi_L^\dagger\nonumber\\
\Ctilde: &  ~~~~  \psi_R \to \etatilde_c\psi_R^\dagger  ,  ~~~~~\psi_L \to \etatilde_c \psi_L^\dagger.
\end{align}
Hence for the AZ classes A,C,D which do not have $\bfT$ or $\bfP$ symmetry, chiral states with only $\psi_R$ or $\psi_L$ can be realized as edge states and  are protected from a mass gap since mass term couples  left and right.

We now use the $\bfT$ symmetry to assign (pseudo-) spins and check for spin-momentum locking. On physical grounds,  we 
 consider the smallest number of components in each class, i.e. either 1 or 2.    
It is well-known that $\bfT$ has the representation $\bfT = \(  \begin{smallmatrix}  0&1 \cr -1&0 \cr \end{smallmatrix} \) $ on spin $1/2$ particles and $\bfT^2=-1$. 
Hence when the representation of $\bfT$ involves $i\sigma_y$ or $i\tau_y$ and $\bfT^2=-1$ in Table ~\ref{tab:25-17}, $\vec{\sigma}$ or $\vec{\tau}$ should act on the spin space. This is particularly interesting since $|\sigma_x=+\rangle$ and $|\sigma_x=-\rangle$ are right- and left-moving states by definition of the hamiltonian Eq.~\eqref{Hform}: this, as we mentioned earlier, is a manifestation of spin-momentum locking.

The classes with spin-momentum locking are {AII$_{(1)}$,    DIII$_{(1)}$, CII$_{(1)}$}. These are  all TI-TS edge states of type $\Z_2$ within our scheme.  For these, we can label the fields $\psi_R = \psi_{R\up} , ~ \psi_L = \psi_{L\down}$.  {AII}$_{(1)}$ and {DIII}$_{(1)}$ have well known examples. QSH edge states\cite{Konig02112007,Bernevig15122006, Kane2005} {\it in the absence of particle hole symmetry} are examples of {AII}$_{(1)}$ class. Note that we  derived here the spin-momentum locking, which arises from the spin-orbit coupling in QSH systems, on very general grounds. 
A 2d version of a He$_3$B superfluid phase where up-spin pairs and down-spin pairs have  opposite angular momentum, would be an example of the            {DIII}$_{(1)}$ class. \footnote{The spin-momentum locking requires equal-spin pairing with {\bf d}-vector in plane\cite{RevModPhys.63.239,Schnyder2008}}. Such a state has not been realized yet, but perhaps could be in a film geometry with  control over the  boundary conditions. 
  {CII}$_{(1)}$ can be realized~\cite{Hsu-Kim2012} as a particle-hole doubled version of  {AII}$_{(1)}$ much the same way as how in 
   3d a  CII TI was constructed out of two copies of 3d Dirac Hamiltonian in \textcite{Schnyder2008}.

{DIII}$_{(2)}$ and {CII}$_{(2)}$ classes have both spin components for right-movers and left-movers each. The  Kane-Mele 
(KM) model\cite{Kane2005} at zero chemical potential  has particle-hole symmetry and hence does not strictly speaking  belong to class AII. Moreover the  spin or charge edge current is  absent as the current operators are odd under charge conjugation~\cite{Zheng2011}. Nevertheless, there is a charge neutral gapless edge mode\cite{Zheng2011,Hohenadler2011a}. This is an example of {DIII}$_{(2)}$ class~\cite{Hsu-Kim2012}.  {CII}$_{(2)}$ is unique in that spin is tied to charge, i.e. particle-hole transformations flip spin: $(\psi_{R\up},  \psi_{R\down}  )  \to (- \psi_{R\down}^\dagger,  \psi_{R\up}^\dagger )$. Note that these  spin-momentum locking properties offer concrete distinctions  between classes ({DIII}$_{(1)}$, {CII}$_{(1)}$)  and ({DIII}$_{(2)}$, {CII}$_{(2)}$).

{AIII}$_{(1)}$, non-chiral {D}, and {BDI}$_{(2)}$ are spinless fermions. Note that we find  the non-chiral {D}  TI
to be  of $\Z_2$ type and distinct from  the chiral ${D}$ which is of $\Z$ type. 

\section{Variations of Luttinger liquids}
We are now in the position to consider how interactions consistent with
the $\bfT, \bfC, \bfP$ symmetries could  affect the $\dbar=1$ edge states. 
In general,  bulk interactions should lead to interactions on the edge. If the bulk stays gapped, one can focus on the edge states even in the presence of interactions. 
While the topological invariants based on single particle wave functions cannot be applied to interacting systems,  the edge state theory can  incorporate the  effects of interactions. 

The fractional quantum Hall  effect (FQH)  is the prime example. 
The FQH edge state resulting from Coulomb interaction in the bulk  has no topological invariant associated with it, 
 while the  integer QHE is associated with  the Chern number\cite{Chern}. However the fractional quantum Hall  edge states are chiral Luttinger liquids  which are related to  the integer quantum Hall  edge states (chiral Fermi liquid) by the  addition of  an exactly marginal perturbation to the Dirac  action\cite{Wen}.
  An exactly marginal perturbation on a non-interacting edge state preserves the gaplessness,
but deforms it into an interacting theory with non-trivial exponents,  fractional charges,  etc. 

Motivated by the FQH case, we classify  the exactly marginal perturbations for each proposed TI-TS's in Table~\ref{tab:non-chiral}, as a way of characterizing the effect of bulk interactions.

The starting point is the action for the generic free Dirac Hamiltonian Eq.~\eqref{2nd}:
\begin{align}
\label{action}
S &\!=\!  \int \!\!dx dt ~ \left[ \psi_R^\dagger ( \d_z + A_x +V_+)   \psi_R \right. \\
 &+  \psi_L^\dagger (  \d_\zbar -  A_x +V_+)   \psi_L \nonumber\\
&+\left.\left(  \psi_L^\dagger (V_- + i A_y)  \psi_R   + h.c. \right)\right]. \nonumber
\end{align}
Recall that $\psi_R$ and $\psi_L$ are vectors in the space represented by $\tau$. $V_+$ can be interpreted as a chemical potential,  or equivalently
the time component of a gauge field as it couples to currents $\psi_R^\dagger V_+ \psi_R +  \psi_L^\dagger V_+  \psi_L$. We set it to zero. 
If $V_- + i A_y $ is one dimensional,  it simply corresponds to a complex mass.  Hence removing
$A_y$ through a unitary transform  $U_\theta$   is equivalent to removing the phase of  the mass by  redefining $\psi_L$. After removing $A_y$,  and absorbing the physical gauge field $A_x$ to the definition of the $\psi$  fields, the action  for the massless zero mode simplifies to 
\begin{equation}
S=\int dx dt \left(\psi^\dagger_R\partial_z\psi_R+\psi_L^\dagger\partial_{\bar z}\psi_L\right).
\end{equation}

We consider left-right current-current perturbations in analogy with Luttinger liquids and single out those preserving the {\bf T}, {\bf C}, {\bf P} of the free theory.
Consider the  currents   $J_L^a  = \psi_L^\dagger t^a  \psi_L$,  $J_R^a  =  \psi_R^\dagger t^a  \psi_R $,
where $t^a$ is a hermitian matrix acting on the $\tau$ space,  
and  define the operator $\CO^a =  J_L^a J_R^a$ (no sum on $a$).  
Since $\psi$ has scaling dimension $1/2$,   the operator $\CO^a $ has dimension two,  i.e.  it  is marginal, 
and a term $g\,   \CO^a$  can be added to the lagrangian.   
For the $T, \Tt, P, \Pt$ symmetries, $\CO^a$ is invariant  if the appropriate $\eta$ commutes with $t^a$.  
For the $C, \Ct$ symmetries which transform  fields into their conjugates, invariance of the operator additionally requires $(t^a)^T = \pm t^a$. 
The renormalization group beta function for $\CO^a$ is in general proportional to  the quadratic Casimir
for the Lie algebra generated by the $t^a$. If this beta function vanishes for a symmetry invariant $\CO^a$, it is an exactly marginal perturbation.  

For all TI-TS's,  the marginal perturbation $\CO^a$ is invariant for $t^a =\one$, 
and we can consider the action
\beq
\label{action2}
S =  \int dx dt ~ \(   \psi_R^\dagger \d_z \psi_R   +  \psi_L^\dagger   \d_\zbar \psi_L   +  g  J_L  J_R \).
\eeq   
Since the currents $J_{L,R}$ 
are then $U(1)$ currents,  the beta function vanishes making 
this perturbation exactly marginal. Eq.~\eqref{action2} describes different versions of Luttinger liquids for different classes. 

The choice $t^a = \tau_y$,  which requires at least 2 components for each chirality,  also yields an invariant $\CO^a$ for
 the classes  DIII$_{(2)}$ and CII$_{(1,2)}$. Since this involves a single $t^a$, it again generates a U(1) current and
 the  associated $\CO^a$ is again exactly marginal.

We list each exactly marginal perturbation for  the above TI-TS's: 
\begin{itemize}
\item {\bf AII$_{(1)}$ and DIII$_{(1)}$}. Both are one-component spin-momentum locked classes. The only allowed perturbation is with $t^a=\one$:
\beq
\label{Oa2}
\CO^a = 
\( \psi^\dagger_{L\down}  \psi_{L\down}   \) 
  \( \psi^\dagger_{R\up}  \psi_{R\up}   \). 
\eeq
The so-called helical liquid for interacting QSH edge state\cite{Wu2006} requires such  a perturbation.  Interestingly such a bulk interaction effect on the edge states has  been recently confirmed\cite{Zheng2011, Hohenadler2011, Hohenadler2011a}. 

\item {\bf DIII$_{(2)}$ and  CII$_{(2)}$}. Both are two-component classes which can be perturbed with  $t^a = \one$ and $t^a = \tau_y$.  
 $t^a = \one$ yields the spin-full Luttinger liquid with
\beq
\label{Oa1}
\CO^a =  \( \psi^\dagger_{L\up}  \psi_{L\up}  + \psi^\dagger_{L\down} \psi_{L\down} \) 
  \( \psi^\dagger_{R\up}  \psi_{R\up}  + \psi^\dagger_{R\down} \psi_{R\down} \). 
  \eeq
Whereas $t^a = \tau_y$ turn $J_L^a$ and $J_R^a$ into a spin-singlet currents and 
\beq
\label{Oa4} 
\CO^a = - \( \psi^\dagger_{L\up}  \psi_{L\down}  -  \psi^\dagger_{L\down} \psi_{L\up} \) 
\( \psi^\dagger_{R\up}  \psi_{R\down}  -  \psi^\dagger_{R\down} \psi_{R\up} \). 
  \eeq
These are new types of Luttinger liquids  which we refer to as the ``spin-singlet liquid''.

\item  {\bf AIII$_{(1)}$},  non-chiral $\bf D$ and  {\bf   BDI$_{(2)}$}.   These are spinless fermion classes which can be single component. They can only be perturbed with $t^a=\one$.   

\item  {\bf CII$_{(1)}$}. This has both particle and hole components with spin-momentum locking for each component.  It is a different kind of  Luttinger liquid,  which we refer to as  the ``double helix'',  since the free part is essentially a doubled KM model.      
\beq
\label{Oa3} 
\CO^a =  \( \psi^\dagger_{L\down}  \psi_{L\down}  + \psi'^\dagger_{L\down} \psi'_{L\down} \) 
  \( \psi^\dagger_{R\up}  \psi_{R\up}  + \psi'^\dagger_{R\up} \psi'_{R\up} \) 
  \eeq

\end{itemize}

Next consider adding more than one perturbation,  i.e. $ \sum_a  g_a \CO^a$. 
In general,  the operator product expansion of $\CO^a$ with $\CO^b$ generates
another $\CO$ operator associated with the current corresponding to $[t^a , t^b]$,
and this gives rise to a renormalization group beta function proportional to the 
quadratic casimir of the Lie algebra generated by the $t^a$.   
  Only 
classes  DIII$_{(1)}$  and CII$_{(2)}$ have two allowed $\CO^a$ listed above:   $t^a = \one ~ {\rm or} ~ \tau_y$.
However since these $t^a$ commute,  this two parameter perturbation is also exactly marginal.  
In summary, we find  all possible symmetry preserving quartic interactions 
to be exactly marginal, deforming the free Dirac edge theory into an interacting one that preserves the  gaplessness .

\section{Conclusions}

We classified Dirac hamiltonians in one dimension according to the discrete symmetries of 
time-reversal,  particle-hole and chiral symmetry,  and found 17 inequivalent ones.   
Assuming  that two-dimensional  topological insulators (or superconductors)  are realized on their
one dimensional boundary as Dirac fermions,   we found 11 of these  classes that  possessed 
a zero mode which  was protected by the symmetries.       This should be compared
with the classifications based on bulk topological or boundary localization properties 
in \cite{Schnyder2008,Ryu2010,kitaev-periodic},  which 
predict 5 classes in any dimension.      
The classes we find beyond the standard 5
are in classes AIII, BDI,   two versions of CII, a distinct version of DIII and a $\Z_2$ version of D.  
We suggested that  physical realizations for the new TI's in classes  CII$_{(1)}$ and CII$_{(2)}$  could perhaps 
be a doubled Kane-Mele model and trigonally strained graphene respectively.  

 The simplest interpretation of the existence of these new classes of TI in two spatial dimensions is that 
there are theories with boundary zero modes that are not necessarily protected by topology,  and this 
is attributed to the richer structure of the classification of Dirac hamiltonians in 1 dimension.
   On the other hand,  it  remains a  possibility  that the new classes are characterized by some
as yet unknown topological invariants.   

We also studied possible manifestations of bulk interactions as quartic interactions on the boundary
in two dimensions.   For all classes of potential TI's,  we found that all such interactions that preserve 
the discrete symmetries are exactly marginal.   The exact marginality preserves the gaplessness,
but deforms the theory into distinct  variations of  Luttinger liquids.

\section{Acknowledgments}
We   wish to thank Ching-Kai Chiu,  Charlie Kane, Andreas Ludwig, and  Michael Stone  for useful discussions.  In particular, we thank Shinsei Ryu and Hong Yao for sharing their manuscript in preparation, on classification of topological insulators protected by reflection symmetry.  AL  thanks the  Ecole Normale Sup\'erieure and
LPTHE in Paris,  and CBPF in Rio de Janeiro,   for their hospitality and support. E-A.K thanks the Kavali Institute for Theoretical Physics, UCSB, for their hospitality and support. 
This work is supported by the National Science Foundation under grant numbers  PHY-0757868 (AL), 
DMR-0520404 (E-A.K),   NSF 
CAREER  DMR-0955822 (E-A. K)   and by the ``Agence Nationale de la Recherche'' contract ANR-2010-BLANC-0414 (DB).


%

\end{document}